\documentclass[paper,12pt,nofootinbib,notitlepage]{revtex4-1}
\usepackage{graphicx}
\usepackage{amsmath}
\usepackage{amssymb}
\usepackage{dcolumn}
\usepackage{bm}
\usepackage{color}
\usepackage{dsfont}
\usepackage{feynmp}
\usepackage{marvosym}
\usepackage{phaistos}

\newcommand \beq{\begin{eqnarray}}
\newcommand \eeq{\end{eqnarray}}

\begin{document}
\unitlength=1mm
\allowdisplaybreaks

\title{On overlapping Feynman (sub)graphs}

\author{Urko Reinosa}
\affiliation{Centre de Physique Th\'eorique (CPHT), CNRS, Ecole Polytechnique,\\ Institut Polytechnique de Paris,  Route de Saclay, F-91128 Palaiseau, France.}

\date{\today}

\begin{abstract}
We discuss, on general grounds, how two subgraphs of a given Feynman graph can overlap with each other. For this, we use the notion of connecting and returning lines that describe how any subgraph is inserted within the original graph. This, in turn, allows us to derive ``non-overlap'' theorems for one-particle-irreducible subgraphs with $2$, $3$ and $4$ external legs. As an application, we provide a simple justification of the skeleton expansion for vertex functions with more than five legs, in the case of scalar field theories. We also discuss how the skeleton expansion can be extended to other classes of graphs.
\end{abstract}

\maketitle

\section{Introduction}
Overlapping divergences make the practical treatment of UV divergences in a quantum field theory cumbersome. In modern approaches, there exist various ways of tackling this issue, all based, in one way or another, on the use of infinitesimal or finite variations of the Feynman graphs with respect to appropriate parameters. The best known among these approaches is certainly the functional renormalization group \cite{Dupuis:2020fhh}, but there exist other possibilities, such as the one put forward in Ref.~\cite{tHooft:2004bkn}, see also Ref.~\cite{Baker:1976vz,Collins:1984xc} for even older proposals. The benefit of these approaches is that one does not need to worry about possible overlapping divergences, since they are, if any, automatically disentangled. 

There might be situations, however, where one needs to assess the absence of overlapping divergences in a given quantity build out of Feynman graphs. One recent example of this situation is reported in Ref.~\cite{Blaizot:2021ikl} where the overlapping divergences that appear in the two-particle-irreducible (2PI) formalism for the case of a scalar $\varphi^4$ theory are disentangled with the help of the functional renormalization group and classified into divergences of the two-point function, divergences of the four-point function, and divergences of higher derivatives $\delta^n\Phi[G]/\delta G^n$ (with $n\geq 3$) of the so-called Luttinger-Ward functional $\Phi[G]$, a functional of the propagator that enters the definition of the 2PI effective action. At first sight, there seems to be too many independent divergences as compared to the expected ones in scalar $\varphi^4$ theory. However, a careful analysis reveals that the divergences of $\delta^3\Phi/\delta G^3$ (and then also those of subsequent derivatives) are non-overlapping, which, in turn, implies that they are entirely governed by those of the two- and four-point functions. Here we extend and refine the discussion of Ref.~\cite{Blaizot:2021ikl} to describe, on general grounds, how two subgraphs of a given Feynman graph can overlap. This allows us to derive a series of ``non-overlap'' theorems for one-particle-irreducible subgraphs with $2$, $3$ and $4$ external legs. For other interesting works that relate to overlapping divergences, see for instance \cite{Kreimer:1998iv,Connes:2002ui}.

The absence of overlapping divergences is intimately related to the possibility of constructing a {\it skeleton expansion} for a given vertex function with high enough external legs. By this, it is meant that, instead of computing such vertex function by summing all the Feynman graphs it is made of, one can first sum the so-called skeleton graphs in this list, and then, in each skeleton, replace each line by the full propagator, each tree-level trilinear coupling by the full three-point function (if any), and each tree-level quartic coupling by the full four-point function. In this way, one can hide any reference to the bare mass and the bare trilinear and quartic couplings. This is a well known result quoted for instance in Ref.~\cite{tHooft:2004bkn,Lu:1991qr}, with various applications such as for instance in the context of conformal theory \cite{Mack:1973kaa,Petkou:1994ad,Petkou:1995vu,Petkou:1996np,Goncalves:2018nlv}. A proof of this result is however difficult to find in the literature. In this paper, using the non-overlap theorems metioned above, we provide a simple justification of the skeleton expansion for vertex functions with more than five legs, in the case of scalar field theories. We also discuss how the skeleton expansion can be extended to other classes of graphs, in particular to the derivatives of the Luttinger-Ward functional.

Even though the non-overlap theorems apply to any theory, for convenience we consider the simpler framework of a scalar theory. We do not restrict the type of interaction, however, which could be any $\varphi^n$, with $n\geq 3$. In fact, we could consider various of these interactions simultaneously. In Sec.~II, we introduce various definitions. In particular, we define graphs and subgraphs and describe how a subgraph is inserted within a given graph with the help of connecting and returning lines. In Sec.~III, this is used to describe how two subgraphs of a given graph can overlap with each other. In Sec.~IV, we restrict to the case of one-particle-irreducible subgraphs and derive the non-overlap theorems which are then used in Sec.~V to justify the skeleton expansion of vertex functions with more than five legs. We then extend this result to other classes of functions, including the high enough derivatives of the Luttinger-Ward functional.

\section{Graphs and subgraphs}\label{sec:graphs}
 In perturbative calculations, quantities are computed by summing Feynman graphs made of two basic elements: free propagators that are represented graphically as lines, and vertices that are represented by points with a certain number of legs.\footnote{For instance, the $\varphi^n$ interaction vertex has $n$ legs.} We stress that vertex legs are not to be seen as lines, but rather as little anchors on which lines can be attached (or not). In what follows, we introduce more precisely the notion of graph together with some related concepts. In particular, we describe how a subgraph of a graph is inserted within that graph by means of both connecting lines and returning lines. This will then allow us to describe all possible overlaps between subgraphs of a given graph.

\subsection{Graphs}
We define a {\it graph} ${\cal G}$ as any collection of vertices and lines with the property that the two ends of any line of ${\cal G}$ are attached to vertices of ${\cal G}$. We can distinguish two types of vertices within the graph: those whose legs are all connected to lines of ${\cal G}$ are called {\it internal vertices}, while the others are called {\it external (or boundary) vertices.} The legs of external vertices are of two types: legs attached to lines of ${\cal G}$ and legs attached to no line. We call the latter the {\it external legs} of the graph ${\cal G}$ and denote them as $n_{\rm ext}({\cal G})$ in the following. 

We stress that our definition of graph excludes the possibility of lines with one end not attached to a vertex. This is just a convenient choice for the subsequent discussion, and, if needed, we can always attach such free lines to the external legs of a graph. Reciprocally, any graph including such free lines is associated to a unique graph that has no external lines. We also exclude lines which are not connected to any vertex. These are just trivial elements (disconnected from the rest) that can again be added at will when needed. There are no other restrictions for the moment, so the graphs could be one-particle-reducible, unamputated or even disconnected. Restrictions will be considered when appropriate.

\begin{figure}[t]
\begin{center}
\includegraphics[height=0.3\textheight]{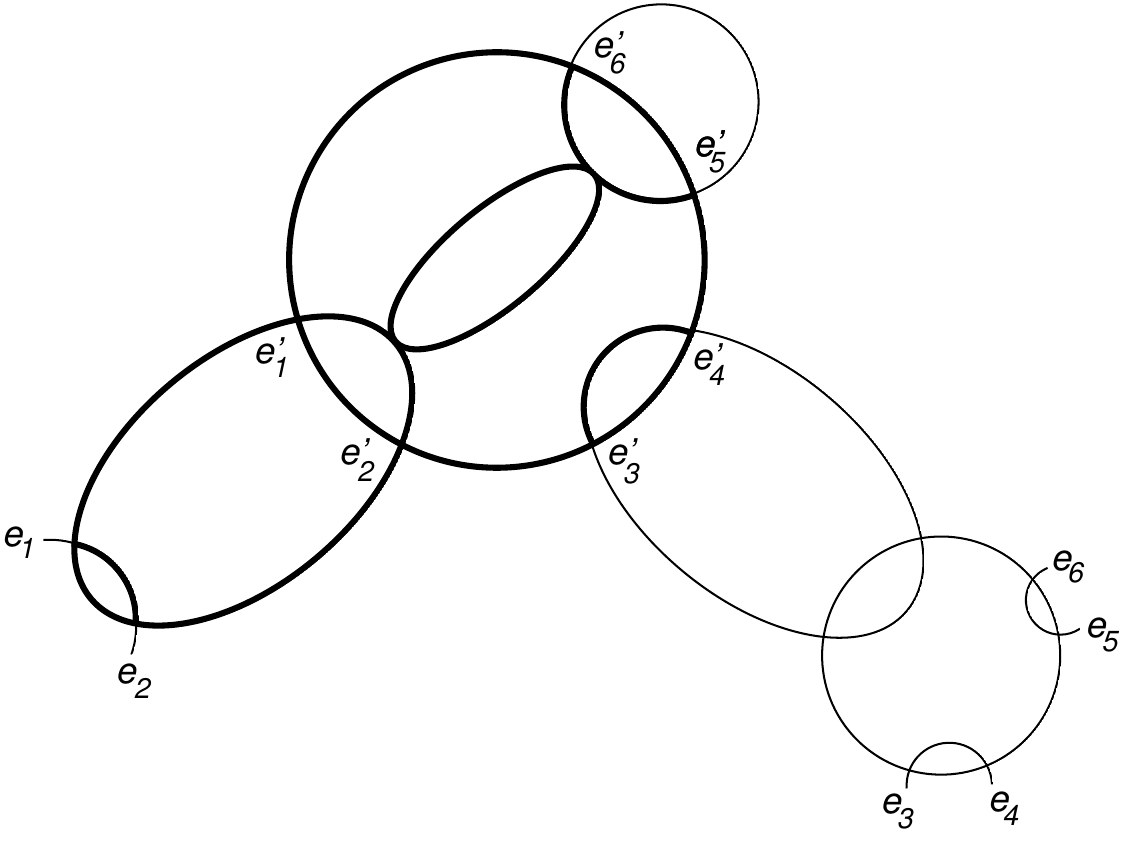}
\caption{An example of graph with six external legs $e_1,\dots, e_6$ in $\varphi^4$ theory. The thick lines highlight one particular subgraph with six external legs as well, $e_1$, $e_2$ and $e'_3,\dots,e'_6$. We have chosen a one-particle-irreducible graph for illustration but the discussion in Secs.~\ref{sec:graphs} and \ref{sec:overlap} applies to any type of graph as defined in Sec.~\ref{sec:graphs}. The leg labels $e'_1$ and $e'_2$ have been introduced for later purpose, see Sec.~\ref{sec:overlap}.}
\label{fig:example}
\end{center}
\end{figure}

In Fig.~\ref{fig:example}, we draw one example of graph in $\varphi^4$ theory. We shall use it recurrently to illustrate the various notions to be introduced below.

\subsection{Subgraphs}
A subgraph $\bar {\cal G}$ of a graph ${\cal G}$ is any collection of vertices and lines of ${\cal G}$ that forms a graph in the sense defined above. We write this as $\bar {\cal G}\subset {\cal G}$.\footnote{In this paper, we use a set theory notation (which differs slightly however from its use in set theory), similar to the one used in \cite{Kreimer:1998iv}.} We mention that any internal vertex of $\bar {\cal G}$ is necessarily an internal vertex of ${\cal G}$. In contrast, an external vertex of $\bar {\cal G}$ can be either an external vertex of ${\cal G}$ or an internal vertex of ${\cal G}$.

Let us also mention that, when seen as a part of ${\cal G}$, the legs of the external vertices of $\bar {\cal G}$ are now of three different types:  legs attached to lines of $\bar {\cal G}$, legs attached to lines of ${\cal G}$ that are not in $\bar {\cal G}$ and legs attached to no line (and thus corresponding to external legs of the original graph ${\cal G}$). Among these three types of legs attached to the external vertices, we refer to the last two as the external legs of the subgraph $\bar {\cal G}$. It is clear that the subgraph can be made a separate entity, disconnected from the original graph, by cutting all lines that are attached to its external legs. Indeed, these are the only lines that connect a vertex of the subgraph to a vertex in the rest of the graph.\footnote{Moreover, once separated from the rest, the external legs of the subgraph $\bar {\cal G}$ coincide with the external legs of the graph $\bar {\cal G}$ seen as a separated entity, as defined in the previous section.}

An example of subgraph is shown in Fig.~\ref{fig:example}. We see clearly what are the external vertices, and thus the external lines that need to be cut to make the subgraph disconnected from the original graph (these are the lines connected to the legs $e'_3,\dots,e'_6$).

\subsection{Connecting lines and returning lines}
The subgraph $\bar {\cal G}$ is said to be {\it dense} within ${\cal G}$ if its vertices exhaust all vertices of ${\cal G}$. In the opposite case, we can define a new subgraph, known as the {\it complementary graph} of $\bar {\cal G}$ within ${\cal G}$, denoted ${\cal G}/\bar {\cal G}$ and formed by all the remaining vertices and all the lines that connect them. Together with the vertices of $\bar {\cal G}$, the vertices of ${\cal G}/\bar {\cal G}$ exhaust all the vertices of ${\cal G}$. This is not so, however, for the lines. Indeed, there might be lines that connect one vertex of $\bar {\cal G}$ and one vertex of ${\cal G}/\bar {\cal G}$ and which, therefore, do not belong neither to $\bar {\cal G}$ nor to ${\cal G}/\bar {\cal G}$. We call these lines the {\it connecting lines} of $\bar {\cal G}$ within ${\cal G}$. Obviously, these can also be seen as the connecting lines of ${\cal G}/\bar{\cal G}$ within ${\cal G}$.

\begin{figure}[t]
\begin{center}
\includegraphics[height=0.38\textheight]{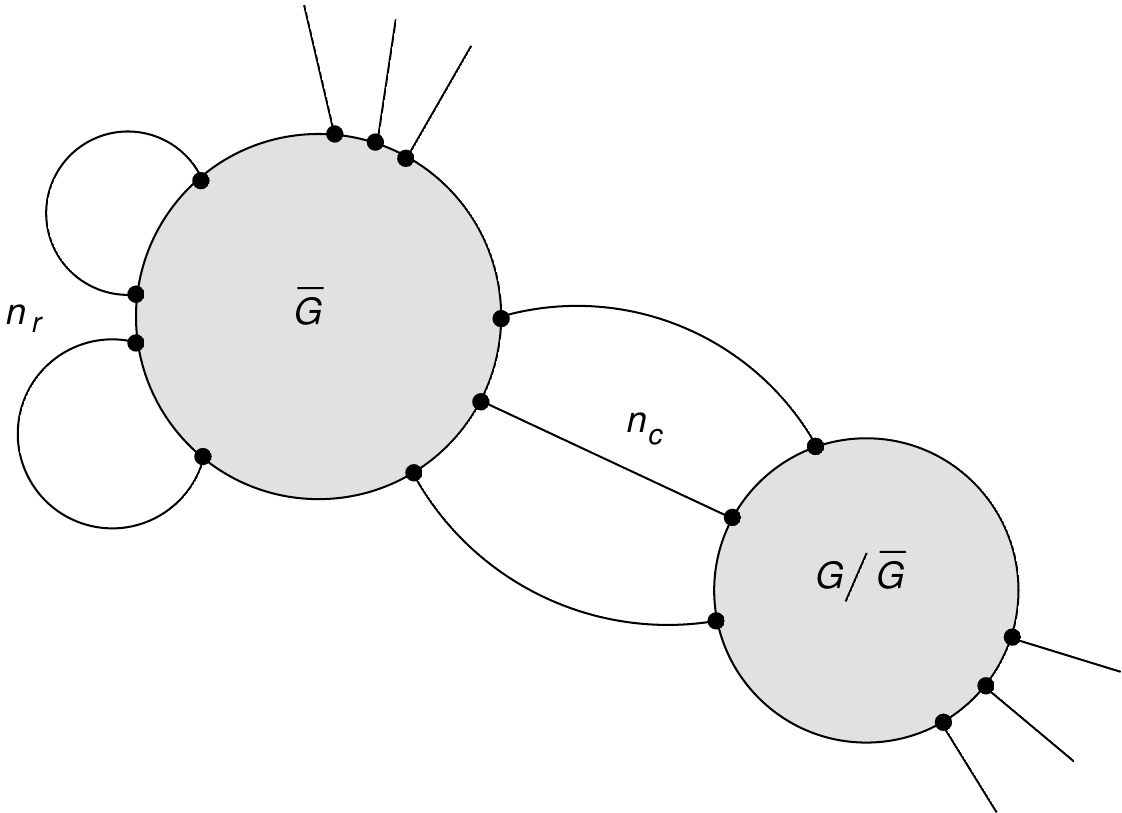}
\caption{A subgraph $\bar {\cal G}$ is inserted in a graph ${\cal G}$ by means of $n_c$ connecting lines and $n_r$ returning lines. By definition, the complementary graph ${\cal G}/\bar {\cal G}$ has no returning lines. In contrast, we cannot hide the returning lines of $\bar {\cal G}$ via a redefinition of $\bar {\cal G}$ because, in general, $\bar {\cal G}$ will be forced upon us by the context.}
\label{fig:subgraph}
\end{center}
\end{figure}

There might also be certain lines that connect vertices of $\bar {\cal G}$ but which do not belong to $\bar {\cal G}$. We call these  {\it returning lines} of $\bar {\cal G}$ within ${\cal G}$. Such lines can exist because, when selecting the subgraph $\bar {\cal G}$, we choose lines and vertices of ${\cal G}$ but our choice does not necessarily include all lines that connect the selected vertices with each other. On the other hand, our definition of the complementary subgraph ${\cal G}/\bar {\cal G}$ is such that all the lines connecting the vertices of ${\cal G}/\bar {\cal G}$ are elements of ${\cal G}/\bar {\cal G}$.  In other words, ${\cal G}/\bar {\cal G}$ does not have any returning lines within ${\cal G}$. Of course, we could redefine the subgraph $\bar {\cal G}$ such that it includes the returning lines as well, and therefore such that $\bar {\cal G}$ and ${\cal G}/\bar {\cal G}$ are treated in a more symmetrical way. However, we shall not do this here for a precise reason: in the following discussion, the subgraph $\bar {\cal G}$ will be imposed on us by the context, while we will always be free to choose ${\cal G}/\bar {\cal G}$ such that it does not have any returning lines.  Note finally that, in the case of a dense subgraph, there are only returning lines, no connecting lines. On the other hand, the absence of connecting lines does not necessarily imply that the subgraph $\bar {\cal G}$ is dense within ${\cal G}$ since the complementary subgraph ${\cal G}/\bar {\cal G}$ could be disconnected from $\bar {\cal G}$. The equivalence works in the case of a connected graph ${\cal G}$ though.

The notions of connecting and returning lines provide a graphical representation of how a given subgraph $\bar {\cal G}$ is inserted within a graph ${\cal G}$, see Fig.~\ref{fig:subgraph}. This structure will be central in the following developments. If we take the example of Fig.~\ref{fig:example}, we see that the considered subgraph has one returning line (the thin line connected to the legs $e'_5$ and $e'_6$) and two connecting lines (the two thin lines attached respectively to the legs $e'_3$ and $e'_4$). The remaining lines and vertices (in the bottom right of the figure) form the complementary graph.

\section{Overlapping subgraphs}\label{sec:overlap}
We are now ready to discuss how two subgraphs $\bar {\cal G}_1$ and $\bar {\cal G}_2$ of a given graph ${\cal G}$ can overlap with each other. In fact, for the moment, the original graph ${\cal G}$ will play no role and we can equally think in terms of the overlap of two original graphs.

By overlapping subgraphs or graphs, we mean that $\bar {\cal G}_1$ and $\bar {\cal G}_2$ have certain vertices and lines in common. We shall in fact consider the collection of all common vertices and lines between $\bar {\cal G}_1$ and $\bar {\cal G}_2$. It is quite obvious that, if a line is common to $\bar {\cal G}_1$ and $\bar {\cal G}_2$, then the two vertices attached to its ends are also common to $\bar {\cal G}_1$ and $\bar {\cal G}_2$. It follows that this common collection of lines and vertices forms a graph, referred to as the {\it overlap graph between $\bar {\cal G}_1$ and $\bar {\cal G}_2$,} which we denote $\bar {\cal G}_1\cap \bar {\cal G}_2$ in what follows.

\begin{figure}[t]
\begin{center}
\includegraphics[height=0.32\textheight]{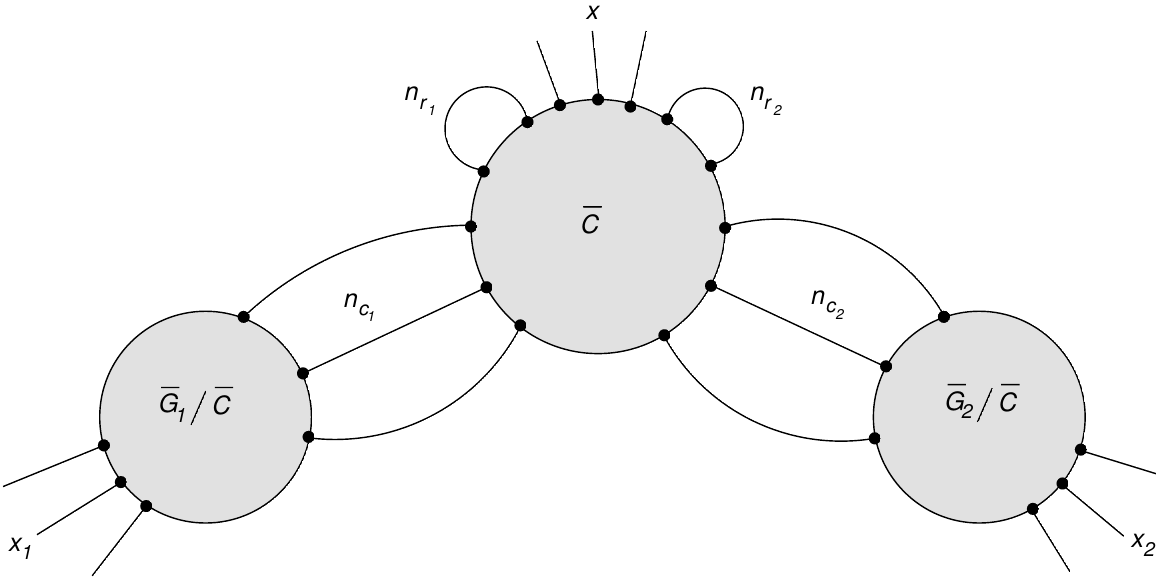}
\caption{Overlap between two graphs $\bar {\cal G}_1$ and $\bar {\cal G}_2$. The overlap graph $\smash{\bar {\cal C}=\bar{\cal G}_1\cap\bar{\cal G}_2}$ is the (maximal) common subgraph of $\bar {\cal G}_1$ and $\bar {\cal G}_2$. This common subgraph has $n_{c_i}$ connecting lines within $\bar {\cal G}_i$ and $n_{r_i}$ returning lines within $\bar {\cal G}_i$. We denote by $x$, $x_1$ and $x_2$ the numbers of external legs of $\bar {\cal C}$, $\bar {\cal G}_1/\bar {\cal C}$ and $\bar {\cal G}_2/\bar {\cal C}$ that are connected neither to connecting lines nor to returning lines.}
\label{fig:overlap}
\end{center}
\end{figure}

\subsection{Overlap pattern}
This common graph $\bar {\cal G}_1\cap \bar {\cal G}_2$ is in fact a subgraph of both $\bar {\cal G}_1$ and $\bar {\cal G}_2$. We can then apply the results of the previous section twice and introduce two sets of connecting lines, $n_{c_1}$ and $n_{c_2}$ in number, as well as two sets of returning lines, $n_{r_1}$ and $n_{r_2}$ in number. This leads to the graphical representation shown in Fig.~\ref{fig:overlap}, where, for later use, we have also introduced the numbers of external legs of $\bar {\cal G}_1\cap \bar {\cal G}_2$, $\bar {\cal G}_1/(\bar {\cal G}_1\cap \bar {\cal G}_2)$ and $\bar {\cal G}_2/(\bar {\cal G}_1\cap \bar {\cal G}_2)$ attached neither to connecting lines nor to returning lines, and denoted respectively as $x$, $x_1$ and $x_2$. It is important to stress that no connecting line of $\bar {\cal G}_1\cap \bar {\cal G}_2$ within $\bar {\cal G}_1$ can be a connecting line of $\bar {\cal G}_1\cap \bar {\cal G}_2$ within $\bar {\cal G}_2$, or vice-versa. Otherwise, this line would be a line of both $\bar {\cal G}_1$ and $\bar {\cal G}_2$ and then an element of $\bar {\cal G}_1\cap \bar {\cal G}_2$, that is not a connecting line. The same remark applies to the returning lines.

\subsection{Counting external legs}
The external legs of $\bar {\cal G}_1$ are those labelled $x_1$, $x$ as well as the $n_{t_2}\equiv n_{c_2}+2n_{r_2}$ legs attached to the connecting and returning lines of $\bar {\cal C}$ within $\bar {\cal G}_2$. Reciprocally, the external legs of $\bar {\cal G}_2$ are those labelled $x_2$, $x$ as well as the $n_{t_1}\equiv n_{c_1}+2n_{r_1}$legs attached to the connecting and returning lines of $\bar {\cal C}$ within $\bar {\cal G}_1$. We can then write
\beq
n_{\rm ext}(\bar {\cal G}_1) & = & x_1+x+n_{t_2}\,,\label{eq:s1}\\
n_{\rm ext}(\bar {\cal G}_2) & = & x_2+x+n_{t_1}\,.\label{eq:s2}
\eeq
On the other hand, the number of external legs of $\bar {\cal G}_1\cap\bar {\cal G}_2$ is
\beq
n_{\rm ext}(\bar {\cal G}_1\cap\bar {\cal G}_2)=x+n_{t_1}+n_{t_2}\,.\label{eq:s3}
\eeq
Finally, it will be convenient to consider the {\it union} of $\bar {\cal G}_1$ and $\bar {\cal G}_2$ obtained by putting together all the vertices and lines of $\bar {\cal G}_1$ and $\bar {\cal G}_2$. This is clearly a graph which we denote $\bar {\cal G}_1\cup \bar {\cal G}_2$. Its number of external legs is given by
\beq
n_{\rm ext}(\bar {\cal G}_1\cup \bar {\cal G}_2)=x_1+x+x_2\,.\label{eq:s4}
\eeq
Using Eqs.~(\ref{eq:s1})-(\ref{eq:s4}), it is then easily checked that
\beq\label{eq:fusion_2}
n_{\rm ext}(\bar {\cal G}_1\cap\bar {\cal G}_2)+n_{\rm ext}(\bar {\cal G}_1\cup\bar {\cal G}_2)=n_{\rm ext}(\bar {\cal G}_1)+n_{\rm ext}(\bar {\cal G}_2)\,.
\eeq
This formula strongly reminds the well known relation between the cardinals of two finite sets $X_1$, $X_2$ and the cardinals of the sets $X_1\cup X_2$ and $X_1\cap X_2$.\footnote{More precisely, one has ${\rm card}(X_1\cap X_2)+{\rm card}\,(X_1\cup X_2)={\rm card}\,X_1+{\rm card}\,X_2\,.$} We stress however that Eq.~(\ref{eq:fusion_2}) is not a trivial application of the corresponding formula between the cardinals of the sets of external legs of $\bar {\cal G}_1$, $\bar {\cal G}_2$, $\bar {\cal G}_1\cup\bar {\cal G}_2$ and $\bar {\cal G}_1\cap \bar {\cal G}_2$ because the sets of external legs of $\bar {\cal G}_1$ or $\bar {\cal G}_2$ are not subsets of the set of external legs of $\bar {\cal G}_1\cup \bar {\cal G}_2$ and so the union of the sets of external legs of $\bar {\cal G}_1$ and $\bar {\cal G}_2$ is not the set of external legs of $\bar {\cal G}_1\cup \bar {\cal G}_2$. Instead, the formula (\ref{eq:fusion_2}) needs to be seen as consequence of the overlapping structure depicted in Fig.~\ref{fig:overlap}.
 
 \subsection{Listing the possible overlaps}\label{sec:list}
The previous formulas allow us to list all possible overlaps between $\bar {\cal G}_1$ and $\bar {\cal G}_2$. First it follows from Eq.~(\ref{eq:fusion_2}) that a necessary condition for $\bar {\cal G}_1$ and $\bar {\cal G}_2$ to have an overlap is that
 \beq\label{eq:central}
n_{\rm ext}(\bar {\cal G}_1\cap \bar {\cal G}_2)\leq n_{\rm ext}(\bar {\cal G}_1)+n_{\rm ext}(\bar {\cal G}_2)\,.
 \eeq 
This also means that, given $n_{\rm ext}(\bar {\cal G}_1)$ and $n_{\rm ext}(\bar {\cal G}_2)$, we can obtain all possible overlaps between $\bar {\cal G}_1$ and $\bar {\cal G}_2$ by considering all possible values of $n_{\rm ext}(\bar {\cal G}_1\cap\bar{\cal G}_2)$ compatible with the constraint (\ref{eq:central}) and, for each of these values, solve the system (\ref{eq:s1})-(\ref{eq:s3}) for $x$, $x_1$ and $x_2$ as a function of $n_{t_1}$ and $n_{t_2}$. One finds
\beq
x & = & n_{\rm ext}(\bar {\cal G}_1\cap\bar {\cal G}_2)-n_{t_1}-n_{t_2}\,,\label{eq:x}\\
x_1 & = & n_{\rm ext}(\bar {\cal G}_1)-n_{\rm ext}(\bar {\cal G}_1\cap\bar {\cal G}_2)+n_{t_1}\,,\label{eq:x1}\\
x_2 & = & n_{\rm ext}(\bar {\cal G}_2)-n_{\rm ext}(\bar {\cal G}_1\cap\bar {\cal G}_2)+n_{t_2}\,,\label{eq:x2}
\eeq
with the constraints
\beq
&& n_{t_1}+n_{t_2}\leq n_{\rm ext}(\bar {\cal G}_1\cap\bar {\cal G}_2)\,,\label{eq:c1}\\
&& n_{t_1}\geq n_{\rm ext}(\bar {\cal G}_1\cap\bar {\cal G}_2)-n_{\rm ext}(\bar {\cal G}_1)\,,\label{eq:c2}\\
&& n_{t_2}\geq n_{\rm ext}(\bar {\cal G}_1\cap\bar {\cal G}_2)-n_{\rm ext}(\bar {\cal G}_2)\,.\label{eq:c3}
\eeq
Any possible solution defined by the values of $x$, $x_1$, $x_2$, $n_{t_1}$ and $n_{t_2}$ is called an {\it overlap mode}. In App.~A, we determine the number of overlap modes and in Table I, we collect the various overlap modes for subgraphs with $0$ and $1$ external legs. An example of overlap of two subgraphs with $6$ external legs each is provided in Fig.~\ref{fig:example} where the highlighted subgraph, obtained by cutting the lines attached to the legs $e'_3,\dots,e'_6$, overlaps with the subgraph obtained by cutting the lines attached to the legs $e'_1$ and $e'_2$. This overlap mode is characterized by $n_{t_1}=4$ (with $n_{c_1}=2$ and $n_{r_1}=1$), $n_{t_2}=2$ (with $n_{c_2}=2$ and $n_{r_2}=0$), $x=0$, $x_1=2$ and $x_2=4$. 

We mention that the above equations make no direct reference to the numbers or connecting lines $n_{c_i}$ or to the numbers of returning lines $n_{r_i}$, but rather to the combination $n_{t_i}=n_{c_i}+2n_{r_i}$ and, in fact, one can interprete the returning lines as a degenerate case of connecting line which does not connect to any complementary graph but loops back instead to the subgraph. This allows to simplify the graphical representation given in Fig.~\ref{fig:overlap} by ignoring the returning lines, or, more precisely, by hiding them as part of the connecting lines.\footnote{We could introduce {\it overlapping submodes} by considering all the possible ways one can distribute the given $n_{t_i}=n_{c_i}+2n_{r_i}$ among $n_{c_i}$ and $n_{r_i}$. However the distinction between connecting lines and returning lines does not play a very deep role in what follows, see however the short discussion in Sec.~\ref{sec:E}.}\\

\begin{table}[ht]
\begin{tabular}{|| c|c|c|c|c|c|c|c|c ||}
\hline
\;$n_{\rm ext}({\bar {\cal G}_1})$\; & \;\;$n_{\rm ext}({\bar {\cal G}_2})$\;\; & \;\;$n_{\rm ext}({\bar {\cal G}_1\cap\bar {\cal G}_2})$\;\; & \;\;$n_{t_1}$\;\; & \;\;$n_{t_2}$\;\; & \;\;$x$\;\; & \;\;$x_1$\;\; & \;\;$x_2$\;\; & \;\;$n_{\rm ext}({\bar {\cal G}_1}\cup\bar {\cal G}_2)$\;\;\\
\hline \hline
\;0\;\; & \;\;0\;\; & \;\;0\;\; & \;\;0\;\; & \;\;0\;\; & \;\;0\;\; & \;\;0\;\; & \;\;0\;\; & \;\;0\;\;\\
\hline
\;0\;\; & \;\;1\;\; & \;\;0\;\; & \;\;0\;\; & \;\;0\;\; & \;\;0\;\; & \;\;0\;\; & \;\;1\;\; & \;\;1\;\;\\
\hline
\;0\;\; & \;\;1\;\; & \;\;1\;\; & \;\;1\;\; & \;\;0\;\; & \;\;0\;\; & \;\;0\;\; & \;\;0\;\; & \;\;0\;\;\\
\hline
\;1\;\; & \;\;1\;\; & \;\;0\;\; & \;\;0\;\; & \;\;0\;\; & \;\;0\;\; & \;\;1\;\; & \;\;1\;\; & \;\;2\;\;\\
\hline
\;1\;\; & \;\;1\;\; & \;\;1\;\; & \;\;0\;\; & \;\;0\;\; & \;\;1\;\; & \;\;0\;\; & \;\;0\;\; & \;\;1\;\;\\
\hline
\;1\;\; & \;\;1\;\; & \;\;1\;\; & \;\;0\;\; & \;\;1\;\; & \;\;0\;\; & \;\;0\;\; & \;\;1\;\; & \;\;1\;\;\\
\hline
\;1\;\; & \;\;1\;\; & \;\;1\;\; & \;\;1\;\; & \;\;0\;\; & \;\;0\;\; & \;\;1\;\; & \;\;0\;\; & \;\;1\;\;\\
\hline
\;1\;\; & \;\;1\;\; & \;\;2\;\; & \;\;1\;\; & \;\;1\;\; & \;\;0\;\; & \;\;0\;\; & \;\;0\;\; & \;\;0\;\;\\
\hline
\end{tabular}
\caption{\label{table1} overlap modes between subgraphs with $0$ and $1$ external legs as obtained using Eqs.~(\ref{eq:x}), (\ref{eq:x1}) and (\ref{eq:x2}) and the constraints (\ref{eq:c1})-(\ref{eq:c3}). We have omitted certain cases that are deduced from the ones in the table using the exchange $\bar {\cal G}_1\leftrightarrow \bar {\cal G}_2$. For the cases listed here $n_{t_i}\leq 1$ and thus there are no returning lines.}
\end{table}

So far the analysis concerned any type of subgraphs of a given graph. In particular, the subgraphs did not need to be connected. In the next section, we particularize the analysis to specific classes of subgraphs for which we show that overlaps are not possible.

\section{The case of one-particle-irreducible subgraphs}
A {\it one-particle-irreducible (1PI) subgraph} is a connected graph that cannot be made into two disconnected pieces by cutting just one line. In this section we analyze  the possibility of overlap between two such subgraphs. More precisely, we look for generic enough conditions under which such overlaps are excluded. Of course, it will be implicitly assumed here that none of the subgraphs in question is a subgraph of the other one (in particular, they are assumed to be distinct). Otherwise they always overlap, in a trivial manner. Moreover, since the union of two overlapping 1PI subgraphs is also 1PI, it is necessarily contained in one of the 1PI components of the original graph ${\cal G}$. Thus, without loss of generality, we can assume that the original graph is 1PI (and in particular connected).

A 1PI subgraph with $p$ external legs is called a {\it $p$-insertion.} This notion includes the (1PI) graph itself if the latter has $p$ external legs, and also any single vertex associated to the $\varphi^p$ interaction if the latter was included in the theory. We shall now introduce certain notions associated to $p$-insertions and then analyze the conditions under which the $2$-, $3$- and $4$-insertions cannot overlap.

\subsection{Definitions}\label{sec:def}
A graph ${\cal G}$ is called a {\it $p$-skeleton} if it contains no other $p$-insertion than the graph itself (if it has $p$ external legs) or those made of a single vertex (if the $\varphi^p$ interaction vertex is part of the model). We mention that a 1PI graph is necessarily a $0$-skeleton (since it is connected) and also a $1$-skeleton (since any non-trivial $1$-insertion would be necessarily connected to the rest of the graph by a line). 
More generally, we call $p_1/p_2$-skeleton, a graph that is both a $p_1$- and a $p_2$-skeleton.\\

Given two $p$-insertions $\bar {\cal G}_1$ and $\bar {\cal G}_2$ of a graph ${\cal G}$, it might occur that one of them is a subgraph of the other, say $\bar {\cal G}_1\subset \bar {\cal G}_2$. This defines a partial ordering over the set of $p$-insertions of the graph ${\cal G}$. As any partial ordering over a finite set, it admits maximal elements, that is elements that are larger than any other element that is ordered with respect to them. In the present context, we refer to these maximal elements as {\it maximal} $p$-insertions.  They correspond to $p$-insertions that are not themselves subgraphs of another $p$-insertion within the graph ${\cal G}$. Obviously, a maximal $p$-insertion cannot be a subgraph of another maximal $p$-insertion of the same graph, unless these two maximal $p$-insertions coincide.\\

The union of two overlapping $p$-insertions is another $q$-insertion. This is because, if there was a way to split the resulting graph by cutting one line, the cut should lie in any of the two original $p$-insertions. But this is impossible since the latter are 1PI by definition. If we now consider the particular case of the union of two overlapping (and distinct) maximal $p$-insertions, then necessarily $q\neq p$ since otherwise the union would have created a new $p$-insertion that is distinct from the original ones and that is larger than any of them, in contradiction with the fact that the latter were both assumed to be maximal. We shall make use of this result below.

\subsection{Non-overlap theorems}
It is not very difficult to see what is the added value of considering 1PI subgraphs. First, the number of connecting lines of $\bar {\cal G}_1\cap \bar {\cal G}_2$ within $\bar {\cal G}_i$ is either $n_{c_i}=0$ or $n_{c_i}\geq 2$, and in the first case, we necessarily have $n_{r_i}\geq 1$, otherwise one subgraph would be included in the other one. It follows that $n_{t_i}\geq 2$. From Eq.~(\ref{eq:s3}) this implies $n_{\rm ext}(\bar {\cal G}_1\cap\bar {\cal G}_2)\geq 4$, and, combining this with Eq.~(\ref{eq:fusion_2}), we arrive at
\beq\label{eq:in4}
4+n_{\rm ext}(\bar {\cal G}_1\cup\bar {\cal G}_2)\leq n_{\rm ext}(\bar {\cal G}_1)+n_{\rm ext}(\bar {\cal G}_2)\,.
\eeq
For not two large values of $n_{\rm ext}(\bar {\cal G}_1)$ and $n_{\rm ext}(\bar {\cal G}_2)$ this is a strong constraint on $n_{\rm ext}(\bar {\cal G}_1\cup\bar {\cal G}_2)$ that will allow us finding certain obstructions to the presence of overlaps.\\

Consider first the case of $2$-insertions, with $\smash{p_1\equiv n_{\rm ext}(\bar {\cal G}_1)=2}$ and $\smash{p_2\equiv n_{\rm ext}(\bar {\cal G}_2)=2}$. From Eq.~(\ref{eq:in4}) it follows that $\smash{n_{\rm ext}(\bar {\cal G}_1\cup\bar {\cal G}_2)=0}$.  Since the original graph ${\cal G}$ is assumed to be 1PI, this means that $\bar {\cal G}_1\cup \bar {\cal G}_2$ is the graph ${\cal G}$ itself, and, therefore, that the latter cannot have any external leg. We have thus arrive at a  first ``non-overlap'' theorem: the only possibility for a overlap of $2$-insertions (self-energies) is within a graph with no external legs. In other words:\\

{\bf Theorem 2:} $2$-insertions cannot have an  overlap within a (1PI) graph with external legs.\\

\noindent{Let us mention that we know exactly how this overlap occurs in the case of a graph with no external legs since $n_{\rm ext}(\bar {\cal G}_1\cap \bar {\cal G}_2)=4$ from Eq.~(\ref{eq:fusion_2}) and therefore $\smash{n_{t_1}=n_{t_2}=2}$ from Eqs.~(\ref{eq:c1})-(\ref{eq:c3}).\\

Consider next $\smash{p_1\equiv n_{\rm ext}(\bar {\cal G}_1)=3}$ and $\smash{p_2\equiv n_{\rm ext}(\bar {\cal G}_2)=3}$. In this case, the inequality (\ref{eq:in4}) leaves room for the cases $n_{\rm ext}(\bar {\cal G}_1\cup\bar {\cal G}_2)=0$, $1$, $2$. We could analyze the various overlap modes using the discussion in the previous section. However, our purpose here is to find conditions for non-overlap. To this purpose, we note that in the cases $n_{\rm ext}(\bar {\cal G}_1\cup\bar {\cal G}_2)=0,1$, we have again $\bar {\cal G}_1\cup\bar {\cal G}_2=\bar {\cal G}$ and therefore these cases can only exist if the original  (1PI) graph ${\cal G}$ has $0$ or $1$ external legs. In the case $n_{\rm ext}(\bar {\cal G}_1\cup\bar {\cal G}_2)=2$, we do not necessarily have $\bar {\cal G}_1\cup\bar {\cal G}_2=\bar {\cal G}$, so this case is possible if the original graph has two external legs or if it contains $2$-insertions. We then arrive at a second non-overlap theorem:\\

{\bf Theorem 3:} $3$-insertions cannot have a overlap within a (1PI) $2$-skeleton graph that has strictly more than two external legs.\\

Let us finally consider $\smash{p_1\equiv n_{\rm ext}(\bar {\cal G}_1)=4}$ and $\smash{p_2\equiv n_{\rm ext}(\bar {\cal G}_2)=4}$. In this case, the inequality (\ref{eq:in4}) leaves room for the values $n_{\rm ext}(\bar {\cal G}_1\cup\bar {\cal G}_2)=0,\dots,4$. This situation is a bit peculiar because, contrary to the previous cases, the highest possible value of $n_{\rm ext}(\bar {\cal G}_1\cup\bar {\cal G}_2)$ allowing for an overlap, that is $4$, coincides precisely with the number of external legs of the insertions we are probing. So we cannot just get rid of this overlap mode by restricting to $4$-skeleton graphs, for this would mean that there are no $4$-insertions to consider in the first place (aside from the trivial ones). Here is where the notion of maximal $4$-insertions and the result quoted at the end of Sec.~\ref{sec:def} comes in handy. Indeed, suppose that we restrict our analysis to 1PI graphs that are $2$- and $3$-skeletons. Then, what we can show is the following third non-overlap theorem:\\

{\bf Theorem 4:} Maximal $4$-insertions cannot have a overlap within a (1PI) $2/3$-skeleton graph that has strictly more than three external legs.\\

Restricting to $2/3$-skeleton graph with strictly more than $3$ legs immediately gets rids of the cases $n_{\rm ext}(\bar {\cal G}_1\cup\bar {\cal G}_2)=0,\dots,3$. Restricting to maximal $4$-insertions gets rid of the case $n_{\rm ext}(\bar {\cal G}_1\cup\bar {\cal G}_2)=4$ because, as we already discussed above, the union of two overlapping (distinct) maximal $4$-insertions cannot be a $4$-insertion.\\

The above results have been obtained by using the inequality (\ref{eq:in4}). An altenerative strategy consists in listing all possible overlaps of a given type and check that none of them fulfills the premises of the above theorems. This is done in Fig.~\ref{fig:list} where the possible overlaps between $2$-, $3$- and $4$-insertions are listed. In each figure, the blobs make reference to the blobs in Fig.~\ref{fig:overlap}, with the little difference that we have hidden the returning lines as part of the connecting lines, see the discussion at the end of Sec.~\ref{sec:list}.
 
 \begin{figure}[t]
\begin{center}
\hspace{-10.5cm}\includegraphics[height=0.10\textheight]{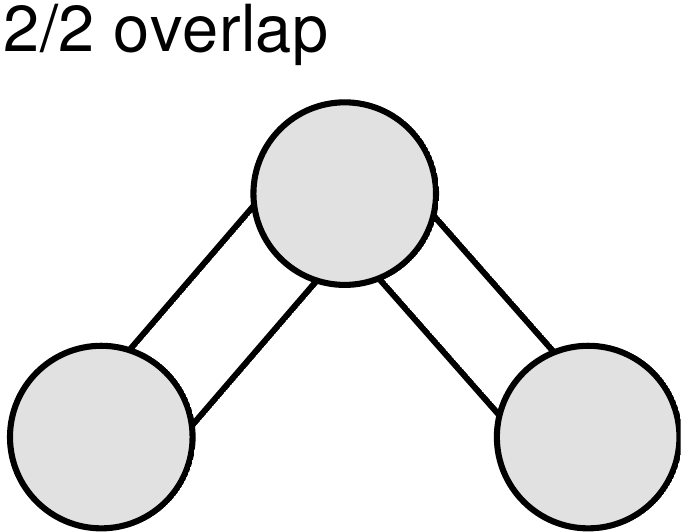}\\

\vglue6mm

\hspace{-6.7cm}\includegraphics[height=0.22\textheight]{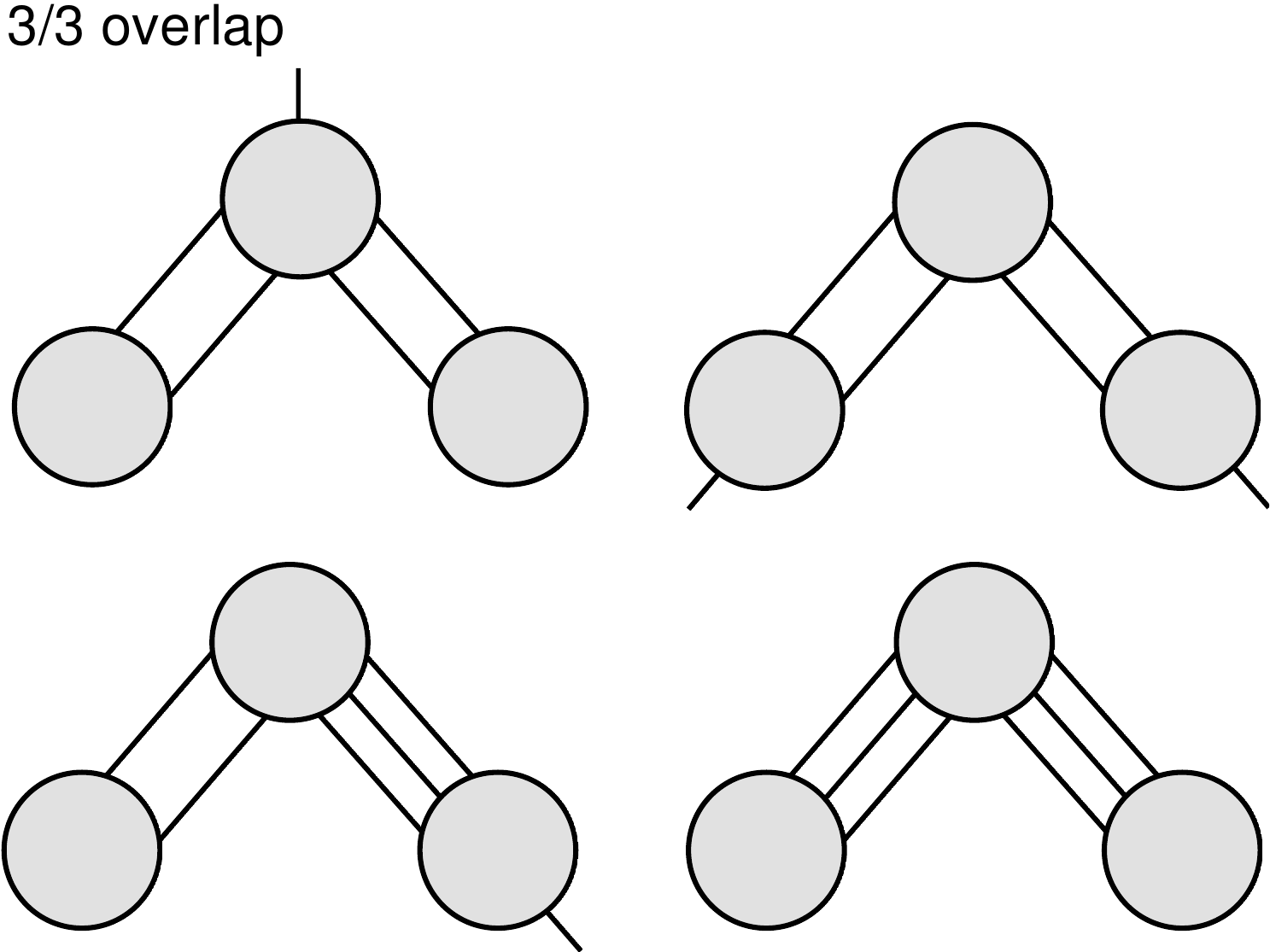}\\

\vglue6mm

\hspace{-3.1cm}\includegraphics[height=0.33\textheight]{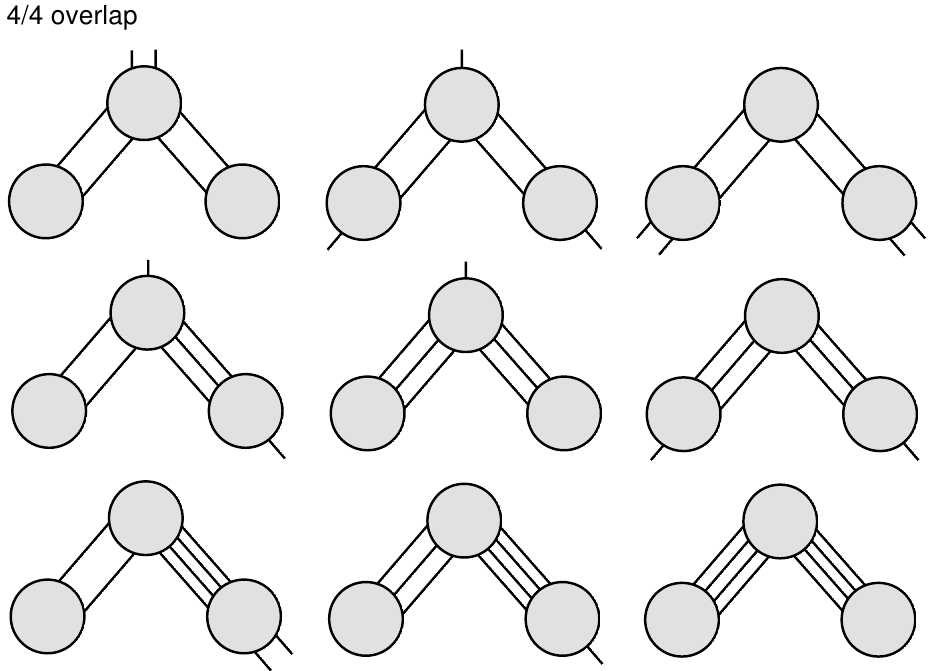}\\

\caption{Possible overlaps between two $2$-insersions, two $3$-insertions or two $4$-insertions. None of them complies with the premises of the theorems $2$, $3$ and $4$. In other words, the theorems apply whenever their premises are fulfilled.}
\label{fig:list}
\end{center}
\end{figure}
 
\subsection{Overlapping insertions of different order}
We can also consider overlaps between insertions of different order. Take for instance $p_1=2$ and $p_2=3$. The inequality (\ref{eq:in4}) implies $n_{\rm ext}(\bar {\cal G}_1\cup\bar {\cal G}_2)\leq 1$. Thus an overlap between these insertions can only occur in graphs with zero or one external leg, or, in other words, a $2$-insertion and a $3$-insertion cannot have an overlap within a graph with strictly more than one leg:\\

{\bf Theorem 23:} A $2$-insertion and a $3$-insertion cannot overlap within a (1PI) graph with strictly more than one external leg.\\

For $\smash{p_1=2}$ and $\smash{p_2=4}$, an overlap can occur only in the cases $n_{\rm ext}(\bar {\cal G}_1\cup\bar {\cal G}_2)=0,1,2$.  This is a situation similar to the one we encountered for the overlap of two $4$-insertions: the highest possible value of $n_{\rm ext}(\bar {\cal G}_1\cup\bar {\cal G}_2)$ allowing for an overlap, that is $2$, coincides precisely with the number of external legs of one of the insertions we are probing. Consider then not an overlap between an arbitrary $2$-insertion and an arbitrary $4$-insertion, but rather between a maximal $2$-insertion and an arbitrary $4$-insertion. This type of overlap cannot occur in a graph with strictly more than one external leg. Indeed, the cases $n_{\rm ext}(\bar {\cal G}_1\cup\bar {\cal G}_2)=0,1$ are trivially excluded, whereas the case $n_{\rm ext}(\bar {\cal G}_1\cup\bar {\cal G}_2)=2$ is excluded because otherwise $\bar {\cal G}_1\cup\bar {\cal G}_2$ would correspond to a $2$-insertion that contains strictly $\bar {\cal G}_1$, in contradiction with the fact that $\bar {\cal G}_1$ was assumed to be maximal. We arrive then at the following result\\

{\bf Theorem 24:} A maximal $2$-insertion and a $4$-insertion cannot overlap within a (1PI) graph with strictly more than one external leg.\\

Finally, for $\smash{p_1=3}$ and $\smash{p_2=4}$, an overlap can occur only in the cases $n_{\rm ext}(\bar {\cal G}_1\cup\bar {\cal G}_2)=0,1,2,3$. This is a situation similar to the one we encountered for the overlap of two $4$-insertions: the highest possible value of $n_{\rm ext}(\bar {\cal G}_1\cup\bar {\cal G}_2)$ allowing for an overlap, that is $3$, coincides precisely with the number of external legs of one of the insertions we are probing. Consider then not an overlap between an arbitrary $3$-insertion and an arbitrary $4$-insertion, but rather between a maximal $3$-insertion and a arbitrary $4$-insertion. This type of overlap cannot occur in a $2$-skeleton graph with strictly more than two external legs. Indeed, the cases $n_{\rm ext}(\bar {\cal G}_1\cup\bar {\cal G}_2)=0,1,2$ are excluded in a trivial way, whereas the case $n_{\rm ext}(\bar {\cal G}_1\cup\bar {\cal G}_2)=3$ is excluded because otherwise $\bar {\cal G}_1\cup\bar {\cal G}_2$ would correspond to a $3$-insertion that contains strictly $\bar {\cal G}_1$, in contradiction with the fact that $\bar {\cal G}_1$ was assumed to be maximal. We arrive then at the following result\\

{\bf Theorem 34:} A maximal $3$-insertion and a $4$-insertion cannot overlap within a (1PI) $2$-skeleton graph with strictly more than two external legs.\\

These results can once again be derived by listing all possible overlaps between $2$-, $3$- and $4$-insertions, see Fig.~\ref{fig:list2}.
 \begin{figure}[t]
\begin{center}
\hspace{-9.6cm}\includegraphics[height=0.11\textheight]{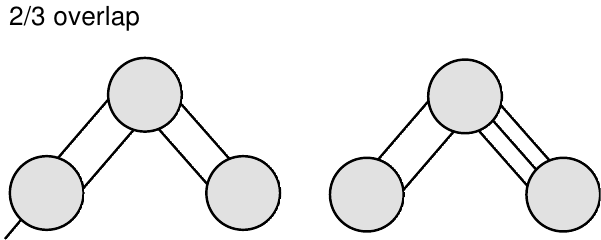}\\

\vglue6mm

\hspace{-6.2cm}\includegraphics[height=0.11\textheight]{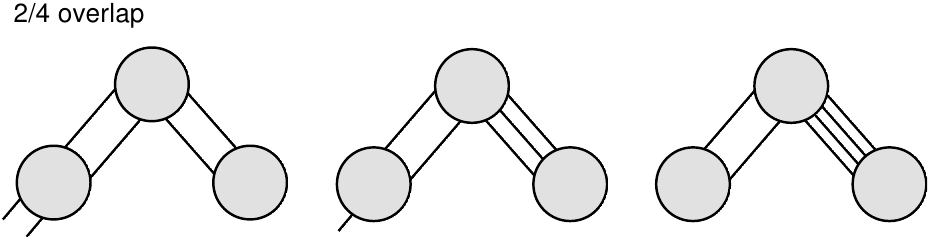}\\

\vglue6mm

\includegraphics[height=0.11\textheight]{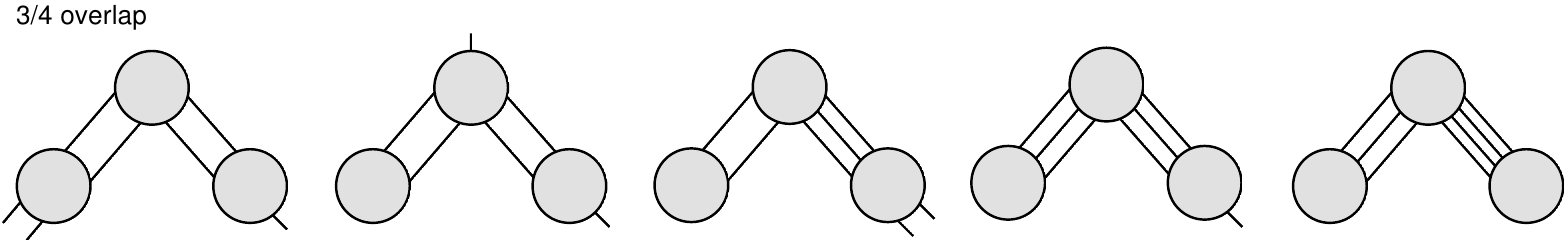}\\

\caption{Possible overlaps between a $2$-insersion and a $3$-insertion,  a $2$-insersion and a $4$-insertion, and a $3$-insersion and a $4$-insertion. None of them complies with the premises of the theorems $23$, $24$ and $34$. In other words, the theorems apply whenever their premises are fulfilled. We note that the first diagram in the last row has strictly more than $3$ external legs, just as in the premise of theorem $34$. However the $3$-insertion that overlaps with the $4$-insertion is not maximal, so this case is not excluded by the theorem.}
\label{fig:list2}
\end{center}
\end{figure}

\subsection{Connecting lines versus returning lines}\label{sec:E}
So far, we made no distinction between connecting and returning lines. This was possible because they play essentially the same role. In particular, with insertions, we have $n_{c_i}\geq 2$ or, in the case where $n_{c_i}=0$, $2n_{r_i}\geq 2$ which allowed us to use $n_{t_i}\geq 2$. One may want to make a distinction between connecting lines and returning lines, and, in particular treat the cases $n_{c_i}=0$ and $n_{c_i}\geq 2$ separately. When proceeding this way, one is lead to discuss three cases of overlap, a {\it generic overlap} with $n_{c_1}\geq 2$ and $n_{c_2}\geq 2$ and {\it non-generic overlaps} with $n_{c_1}=0$ or $n_{c_2}=0$, or both. Using a terminology that we introduced above, the generic overlap corresponds to the case where $\bar {\cal G}_1\cap \bar {\cal G}_2$ is neither dense within $\bar {\cal G}_1$ nor within $\bar {\cal G}_2$, whereas the non-generic overlaps correspond to the cases where $\bar {\cal G}_1\cap\bar {\cal G}_2$ is dense either within $\bar {\cal G}_1$ or within $\bar {\cal G}_2$, or within both.

There is nothing to add to the discussion in the previous section in the case of a generic overlap. In the case of non-generic overlaps, however, the analysis can be slightly refined. Indeed, if $n_{c_1}=0$ and because the subgraphs under consideration are connected, we need to have $x_1=0$ which implies that $n_{\rm ext}(\bar {\cal G}_1\cup\bar {\cal G}_2)=x+x_2$. Moreover, because $n_{r_1}\geq 1$, it follows from Eq.~(\ref{eq:s2}) that $n_{\rm ext}(\bar {\cal G}_1\cup\bar {\cal G}_2)\leq p_2-2$. Similarly, if $n_{c_2}=0$, we need to have $n_{\rm ext}(\bar {\cal G}_1\cup\bar {\cal G}_2)\leq p_1-2$. It is easily seen that these constraints are stronger than (\ref{eq:in4}).

More precisely, in the case $p_1=p_2=2$, we obtain the same constraint $n_{\rm ext}(\bar {\cal G}_1\cup\bar {\cal G}_2)=0$. However, in the case $p_1=p_2=3$, we find $n_{\rm ext}(\bar {\cal G}_1\cup\bar {\cal G}_2)=0,1$ which is the stronger then the constraint that we foud earlier and which allows to enlarge the premise of theorem $3$ to the case of graphs with strictly more than one external leg. Similarly, for $p_1=p_2=4$, we find $n_{\rm ext}(\bar {\cal G}_1\cup\bar {\cal G}_2)=0,1,2$ which allows to enlarge the premise of theorem $4$ to graphs with strictly more than two external legs and to any type of $4$-insertion, not necessarily maximal.\footnote{Arbitrary $4$-insertions can have a generic overlap though. In a $2/3$-skeleton graph, we necessarily have $n_{\rm ext}({\bar {\cal G}_1}\cup {\bar {\cal G}_2})=4$ and then $n_{\rm ext}({\bar {\cal G}_1}\cap {\bar {\cal G}_2})=4$, which, according to (\ref{eq:c1})-(\ref{eq:c3}), leads to all the overlapping modes complying with $n_{t_1}+n_{t_2}\leq 4$. Since $n_{t_i}\geq 2$ in the generic case, we must have $n_{t_i}=2$ and thus $n_{c_i}=2$ and $n_{r_i}=0$.} In the case where $p_1\neq p_2$, with $p_i=2$, $3$ or $4$, it is easily checked that the constraints are the same as those obtained above so the premises of theorems 23, 24 and 34 are not changed.

\subsection{Higher order insertions}
Consider now the case $p_1=p_2=5$. The inequality (\ref{eq:in4}) imposes $n_{\rm ext}(\bar {\cal G}_1\cup\bar {\cal G}_2)\leq 6$. We see here that, even if we restricted to $2/3/4$-skeleton graphs and to maximal $5$-insertions, we could find overlaps with $n_{\rm ext}(\bar {\cal G}_1\cup\bar {\cal G}_2)=6$. From (\ref{eq:fusion_2}), this implies $n_{\rm ext}(\bar {\cal G}_1\cap\bar {\cal G}_2)=4$ which, according to (\ref{eq:c1})-(\ref{eq:c3}), leads to all the overlap modes complying with $n_{t_1}+n_{t_2}\leq 4$. It follows that, without further restrictions, the non-overlap theorems derived above are specific to the cases $p=2$, $3$ and $4$.

We can find nonetheless a non-overlap theorem in the case $p=5$ if we further restrict the possible subgraphs under consideration. Assume for instance that we inquire about the overlap of two-particle-irreducible (2PI) subgraphs, that is subgraphs that cannot be split apart by cutting two lines. In the case of a generic overlap, we have $n_{t_i}\geq 3$ and therefore $n_{\rm ext}(\bar {\cal G}_1\cap\bar {\cal G}_2)\geq 6$, from which it follows now that
\beq
6+n_{\rm ext}(\bar {\cal G}_1\cup\bar {\cal G}_2)\leq n_{\rm ext}(\bar {\cal G}_1)+n_{\rm ext}(\bar {\cal G}_2)\,.
\eeq
This time, overlap of $3$-insertions can only occur if the original graph has no external legs, $4$-insertions cannot overlap if the graph has strictly more than $2$ legs and $5$-insertions cannot overlap if the graph has strictly more than $4$ legs. Moreover, maximal $6$-insertions cannot overlap if the graph has strictly more than $6$ legs.  These results extend almost identically to the case of a non-generic overlap.\footnote{For the $3$-insertions, we need to require the graph to have strictly more than $1$ leg, for the $5$-insertions, it is enough to require the graph to have strictly more than $3$ legs, and for the $6$-inserstions it is enough to require the graph to have strictly more than $4$ legs and the result applies to arbitrary $6$-insertions, not necessarily maximal.}

\section{Application: Hiding the bare mass as well as the trilinear and quartic bare couplings}
In this section we build upon the previous results to show how, for a large class of functions, it is possible to hide the dependence on the bare mass as well as the dependence on the trilinear and quartic bare couplings, using the two-, three- and four-point functions. For simplicity, we first show how this is done for the vertex functions $\Gamma^{(n)}$ with $n\geq 5$. To do so, we show that these vertex functions admit a skeleton expansion, that is rather than computing them by adding all the perturbative graphs they are made of, we can alternatively sum over all the $2/3/4$-skeleton graphs in this list, and then replace each free propagator by the full two-point function $G=[\Gamma^{(2)}]^{-1}$, each tree-level $3$-vertex by the full three point function $\Gamma^{(3)}$ and each tree-level $4$-vertex by the full four-point function $\Gamma^{(4)}$. As already mentioned in the Introduction, this is a known result. It is however interesting to see how it derives from the non-overlap theorems of the previous section. At the end of the section, we argue that this result extends in fact to a larger class of functions.

\subsection{Hiding the bare mass}
Consider a 1PI graph ${\cal G}$ with external legs. We define a {\it chain} of ${\cal G}$ as any connected sequence of lines $G_0$ and $2$-insertions $\Sigma_i$ of the form
\beq
G_0 \Sigma_1 G_0\Sigma_2 G_0\cdots G_0\Sigma_n G_0\,.
\eeq
Since we have assumed that the 1PI graph ${\cal G}$ has external legs, the starting $G_0$ is necessarily different from the ending one. Moreover, we request that this sequence is complete, that is that one cannot add additional $\Sigma_k G_0$'s or $G_0\Sigma_k$'s. In a graph with external legs, it is always possible to identify unambiguously all the chains. It may happen that certain lines $G_0$ are not connected to any self-energy. We call these {\it trivial} chains. 

Given two chains ${\cal C}_1$ and ${\cal C}_2$ of ${\cal G}$, we say that ${\cal C}_2$ is a {\it subchain} of ${\cal C}_1$ if it is a chain of one of the $2$-insertions of ${\cal C}_1$. This relation which we denote as ${\cal C}_2\subset {\cal C}_1$ defines a partial ordering over the set of chains of ${\cal G}$. As any partial ordering over a finite set, it admits maximal elements which we call {\it maximal chains.} Now, according to theorem $2$ above, in a graph with external legs, $2$-insertions cannot have any overlap (unless of course one of them is a subgraph of the other). It is then easily verified that maximal chains cannot have any overlap either (unless of course one of them is a subchain of the other). Let us now use this result to show how to hide the bare mass in $\Gamma^{(n)}$, with $n\geq 3$.

We start by writing $\Gamma^{(n\geq 3)}$ as
\beq
\Gamma^{(n\geq 3)}=\sum_{{\cal D}} {\cal D}[G_0,\{g^{(m\geq 3)}_{\rm b}\}]\,,\label{eq:g3}
\eeq
where the sum runs over all Feynman graphs ${\cal D}$ that contribute to $\Gamma^{(n\geq 3)}$. Depending on the context, ${\cal D}$ denotes the graph itself or the corresponding Feynman integral. It depends on the bare mass $m_{\rm b}$ via the bare free propagator $G_0$. We have also made explicit the dependence on the various bare couplings $\{g^{(m\geq 3)}_{\rm b}\}$.

Since maximal chains do not overlap in the graphs contributing to $\Gamma^{(n\geq 3)}$, one can unambiguously associate to each graph ${\cal D}$, a $2$-skeleton graph denoted ${\cal D}_2$ and obtained from ${\cal D}$ by replacing any maximal chain by a trivial chain. It is convenient to momentarily associate a different label to each trivial chain appearing in ${\cal D}_2$, so that ${\cal D}_2$ is a function of these various chains ${\cal D}_2[G_1,\dots ,G_p,\{g^{(m\geq 3)}_{\rm b}\}]$. The original graph ${\cal D}$ can now be written in terms of its associated $2$-skeleton as
\beq
{\cal D}[G_0,\{g^{(m\geq 3)}_{\rm b}\}]={\cal D}_2[{\cal C}_1,\dots,{\cal C}_p,\{g^{(m\geq 3)}_{\rm b}\}]\label{eq:rel2}
\eeq
where the ${\cal C}_i$ are the maximal chains of ${\cal D}$ that were replaced by trivial chains $G_i$ in order to obtain the $2$-skeleton ${\cal D}_2[G_1,\dots,G_p,\{g^{(m\geq 3)}_{\rm b}\}]$.  We mention that there is no pre-factor in the right-hand side of Eq.~(\ref{eq:rel2}). This is because the symmetry factors factorize: the symmetry factor of ${\cal D}$ equals the symmetry factor of ${\cal D}_2$ times the symmetry factors of the  ${\cal C}_i$. This property relates to the fact that the replacement of maximal chains by trivial chains is unambiguous, see below for more details.

Let now sum both sides of Eq.~(\ref{eq:rel2}) over the graphs ${\cal D}$ contributing to $\Gamma^{(n\geq 3)}$. We perform the sum in two steps. First we sum over all graphs ${\cal D}$ that have the same ${\cal D}_2$, and then we sum over all the possible $2$-skeletons ${\cal D}_2$. Because we are summing over all possible graphs of  $\Gamma^{(n\geq 3)}$ and because this does not put any restrictions on the chains that can appear in the right-hand side of Eq.~(\ref{eq:rel2}) for a given ${\cal D}_2$, we find that the sum over all graphs ${\cal D}$ that share the same ${\cal D}_2$ replaces each chain in the right-hand side of Eq.~(\ref{eq:rel2}) by the sum of all possible chains, that is the two-point function $[\Gamma^{(2)}]^{-1}$:
\beq
{\cal D}_2\big[[\Gamma^{(2)}]^{-1},\dots,[\Gamma^{(2)}]^{-1},\{g^{(m\geq 3)}_{\rm b}\}\big]
\eeq
which we denote for simplicity as ${\cal D}_2\big[[\Gamma^{(2)}]^{-1},\{g^{(m\geq 3)}_{\rm b}\}\big]$. We now need to sum over all possible skeletons ${\cal D}_2$, and we then find
\beq
\Gamma^{(n\geq 3)}=\sum_{{\cal D}_2} {\cal D}_2\big[[\Gamma^{(2)}]^{-1},\{g^{(m\geq 3)}_{\rm b}\}\big]\,.\label{eq:2res}
\eeq

\subsection{Hiding the trilinear and quartic bare couplings}\label{sec:hiding}
Let us now consider $n\geq 4$ and start from Eq.~(\ref{eq:2res}). Since this sum is made of $2$-skeletons that have strictly more than $2$ legs, we can apply theorem $3$. This means that to any graph ${\cal D}_2$, we can unambiguously associate a $2/3$-skeleton ${\cal D}_{23}\in{\cal P}_{23}$ by shrinking any maximal $3$-insertion to a trivial one. Using the same argument as above, we find\footnote{One could wonder here why we are not applying our strategy to $\Gamma^{(3)}$ since theorem $3$ applies to graphs with strictly more than two external legs. This has to do with our choice of definition of a $p$-skeleton diagram which allows for the presence of $p$-insertions equal to the whole graph (in the case the latter has $p$-external legs). Shrinking such insertion to a trivial one would lead to a skeleton graph but would miss many others. We could redefine $p$-skeletons as graphs that contain only trivial, tree-level $p$-insertions. In this case, no skeleton would be missed and Eq.~(\ref{eq:3res}) would apply also to $\Gamma^{(3)}$. This has limited interest however, for the latter identity is a tautology. Similar remarks apply to Eq.~(\ref{eq:4res}) and $\Gamma^{(4)}$.}
\beq
\Gamma^{(n\geq 4)}=\sum_{{\cal D}_{23}} {\cal D}_{23}\big[[\Gamma^{(2)}]^{-1},\Gamma^{(3)},\{g^{(m\geq 4)}_{\rm b}\}\big].\label{eq:3res}
\eeq
Finally, let us now consider $n\geq 5$ and start from Eq.~(\ref{eq:3res}). Since this sum is made of $2/3$-skeletons that have strictly more than $3$ legs, we can apply theorem $4$. Using the same strategy as above, we conclude that
\beq
\Gamma^{(n\geq 5)}=\sum_{{\cal D}_{234}} {\cal D}_{234}\big[[\Gamma^{(2)}]^{-1},\Gamma^{(3)},\Gamma^{(4)},\{g^{(m\geq 5)}_{\rm b}\}\big]\,\label{eq:4res}
\eeq
where the sum runs over the $2/3/4$-skeleton graphs contributing to $\Gamma^{(n\geq 5)}$. Note that it was important to first resum the three-point function. Otherwise, the graphs would not have been $3$-skeletons and we could not have applied theorem $4$. 

One could wonder whether the $2/3$-skeleton graph ${\cal D}_{23}$ (obtained from ${\cal D}$ by first replacing each maximal chain by a trivial chain and then any maximal $3$-insertion by a trivial one) coincides with ${\cal D}_{32}$ (obtained via a similar procedure but in opposite order). The identification ${\cal D}_{23}$=${\cal D}_{32}$ relies on the non-overlap theorem $23$ and grants also that the graphs ${\cal D}_{23}$ are all the $2/3$-skeletons originally present in the collection of graphs ${\cal D}$, that is that no $3$-skeleton graph disappeared in the reduction from ${\cal D}$ to ${\cal D}_2$. Similar remarks apply to ${\cal D}_{234}$, ${\cal D}_{243}$, ${\cal D}_{342}$, ${\cal D}_{324}$, ${\cal D}_{423}$ and ${\cal D}_{432}$. 

Note finally that we cannot continue the procedure (\ref{eq:g3}) $\rightarrow$ (\ref{eq:2res}) $\rightarrow$ (\ref{eq:3res}) $\rightarrow$ (\ref{eq:4res})   further because maximal $5$-insertions can overlap within $2/3/4$-skeleton graphs, as discussed in the previous section.

\subsection{Symmetry factors}
The above results rely on the factorization of symmetry factors. Let us now show how this factorization comes about. Consider for instance a graph ${\cal D}_2\big[[\Gamma^{(2)}]^{-1},\{g^{(m\geq 3)}_{\rm b}\}\big]$. After identifying the maximal $3$-insertions of the graph, which we write $V^{(3)}_1,\dots,V^{(3)}_p$, the graph rewrites in terms of the associated $2/3$-skeleton as
\beq
{\cal D}_2\big[[\Gamma^{(2)}]^{-1},\{g^{(m\geq 3)}_{\rm b}\}\big]=\alpha \,{\cal D}_{23}\big[[\Gamma^{(2)}]^{-1},V^{(3)}_1,\dots,V^{(3)}_3,\{g^{(m\geq 4)}_{\rm b}\}\big]\,,
\eeq
where the pre-factor $\alpha$ accounts for a potential mismatch between the symmetry factor of ${\cal D}_2$ and the symmetry factor of ${\cal D}_{23}$ multiplied by the symmetry factors of the $V^{(3)}_i$. We now show that $\alpha=1$, meaning that the symmetry factors factorize.

To see this, we note that in order to compute the symmetry factor of ${\cal D}_2\big[[\Gamma^{(2)}]^{-1},\{g^{(m\geq 3)}_{\rm b}\}\big]$, we can first compute the symmetry factor of a $(n+3p)$-point function $R$ obtained by chopping off the $3$-insertions from the original graph and then connecting the chopped legs back to the $V^{(3)}_i$. The only thing  that one needs to pay attention to is that, by computing the symmetry factor in this alternative way, we are missing some Wick contraction resulting from the possibility of redistributing the various tree-level vertices of the of the original graph (for simplicity, we assume here that there are only trilinear vertices) among each of the $V^{(3)}_i$ or $R$. Denoting $n_i$ the number of vertices in each of the $V_i$ and by $n$ the number of vertices in the $R$, this produces a factor $(n+n_1+\dots+n_p)!/(n!n_1!\cdots n_p!)$ in the counting of Wick contractions. If we denote by $N_X$ the number of Wick contractions of a given contribution $X$, we have then
\beq
N_{{\cal D}_2}=\frac{(n+n_1+\dots+n_p)!}{n!n_1!\cdots n_p!} N_R\,N_{V^{(3)}_1}\cdots N_{V^{(3)}_p}\,.
\eeq 
But the symmetry factor is equal to the number of Wick contractions divided the factorial of the number of vertices and $3!$ (since we are here considering cubic vertices) elevated to the number of vertices. It follows that
\beq
s_{{\cal D}_2} & = & \frac{N_{{\cal D}_2}}{(n+n_1+\dots+n_p)!(3!)^{n+n_1+\dots+n_p}}\nonumber\\
& = & \frac{N_R}{n!(3!)^n} \frac{N_{V_1}}{n_1!(3!)^{n_1}}\cdots \frac{N_{V_p}}{n_p!(3!)^{n_p}}=s_R s_{V^{(3)}_1}\cdots s_{V^{(3)}_p}\,.
\eeq
Applying the same formula to ${\cal D}_{23}$, with $V^{(3)}_i=g^{(3)}_{\rm b}$, we find $s_R=s_{{\cal D}_{23}}$ and thus
\beq
s_{{\cal D}_2}=s_{{\cal D}_{23}} s_{V^{(3)}_1}\cdots s_{V^{(3)}_p}\,.
\eeq
which is the announced factorization of symmetry factors. The same reasoning applies to the resummation of chains or four-point functions.

\subsection{Extension}
So far we have we have considered the case of 1PI graphs. However, it is pretty clear that our results apply to a larger class of graphs. Consider first the non-overlap theorems. They apply to any disconnected graph whose connected pieces fulfill the premises of these theorems. For instance, theorem $4$ applies to any disconnected graph whose connected parts are $2/3$-skeletons with strictly more than three external legs, and so on. 

Next, let us wonder how the possibility to hide bare parameters extends to functions other than the $\Gamma^{(n)}$'s. Consider for instance a quantity given as an infinite sum of $2$-skeleton graphs whose connected pieces have strictly more than two external legs. It is clear that theorems 2, 3 and 23 apply to each of these graphs and one can therefore associate unambiguously $2/3$-skeleton graphs to each of these graphs. If we now assume that the infinite sum of graphs puts no restriction on the $2$- and $3$-insertions that can appear (this is a property that needs to be verified for each infinite class of graphs that one may consider; it is of course obvious for the $\Gamma^{(n)}$'s) then we can proceed as for the $\Gamma^{(n)}$'s and hide the dependence on the bare mass and trilinear bare coupling using the full two- and three-point functions. 

A direct application of this result is the elimination of the bare parameters in the higher derivatives $\delta^n\Phi/\delta G^n$ (with $n\geq 3$) of the Luttinger-Ward functional $\Phi[G]$. This functional is the sum of two-particle-irreducible graphs with no external legs, that is graphs that cannot be split apart by cutting two lines. The derivatives $\delta^n\Phi/\delta G^n$ are also sums of two-particle-irreducible graphs but only with respect to cuts that leave the external legs associated to a given $\delta/\delta G$ on the same side of the cut. It is easily seen that the connected components of any $\delta^n\Phi/\delta G^n$ with $n\geq 3$ obey the premises of theorem $3$ above. Moreover, the two-particule irreductibility puts no constraint on the possible $3$-insertions that can occur.\footnote{It only imposes that the $3$-insertions cannot be attached to two external legs originating from the same derivative $\delta/\delta G$.} One can then follow the same strategy as in Sec.~\ref{sec:hiding} to show that $\delta^n\Phi/\delta G^n$ with ($n\geq 3$) admits a skeleton expansion in terms of the full two- and three-point functions. The corresponding $2/3$-skeletons obey the premises of theorem $4$ and since the two-particule irreductibility puts no constraint on the possible $4$-insertions that can occur, one can proceed one step further and derive a skeleton expansion  in terms of the full two-, three- and four-point functions. As already mentioned in the Introduction, this has been recently put into good use to formulate a finite set of flow equations for $\Phi$-derivable approximations that make no reference to the bare theory, see Ref.~\cite{Blaizot:2021ikl}.

\section{Final remarks}

\subsection{Applications}
The possibility to express $\Gamma^{(n\geq 5)}$ as an infinite sum of $2/3/4$-skeleton graphs with propagators, three- and four-vertices given respectively by $[\Gamma^{(2)}]^{-1}$, $\Gamma^{(3)}$ and $\Gamma^{(4)}$ is a useful tool in order to truncate infinite hierarchies of equations that appear in continuum approaches to Quantum Field Theory, such as the Dyson-Schwinger tower of equations or the functional renormalization group hierarchy. In such frameworks, a given $n$-point function is typically expressed in terms of higher ones, leading to an infinite tower of equations. Now, by moving deep down enough  the hierarchy and by using the present result, it is clear that one can replace the infinite tower of equations by a finite number of them in which the highest $n$-point functions are expressed in terms of lower ones. The hierarchy is thus closed at the price of expressing some of the $n$-point functions as infinite sums of skeleton graphs in terms of the lower $n$-point functions. But since one can truncate this infinite sum of skeleton according to the number of loops of the skeletons, one obtains a systematically improvable scheme in which one only needs to solve a finite number of equations of the hiearchy.

We also mention that in theories where primarily divergent $n$-point functions have at most $n=4$ external legs, this gives a very graphical explanation of why, in a renormalizable theory, higher $n$-point functions (with $n\geq 5$) are finite once the primarily divergent functions have been renormalized. Indeed once written in terms of $2/3/4$-skeletons, there are no other subdivergences in the graph than those of the two-, three- and four-point functions. Moreover, there are no global divergences since $n\geq 5$. In the case of a theory such as $\varphi^6$ in $d=3$ dimensions, which is also renormalizable but features primarly divergent functions with $6$ legs, this graphical explanation does not apply since maximal $6$-point functions can overlap within any graph and therefore there are inevitably overlapping divergences.

\subsection{Connection to $n$PI effective actions}
The present approach pretty much resembles that followed with $n$-particle-irreducible ($n$PI) effective actions \cite{deDominicis:1964uu,Berges:2004pu,Carrington:2010qq}. Let us here emphasize some differences however. In fact, the present approach deals only with quantities for which the bare mass and the trilinear and quartic bare coupling can be hidden into the two-, three- and four-point functions while avoiding graph over-counting. In contrast, the $n$PI framework deals with the sum of vacuum graphs $\ln Z$ for which none of the above non-overlap theorems apply. Indeed, for such graphs, there is no unambiguous way to identify the maximal $2$-, $3$- and $4$-insertions. It is still possible to rewrite this sum of vacuum graphs as a sum of skeletons.\footnote{The notion of skeleton needs to be slightly extended though, as compared to the definition given in the present paper. For instance, a $2$-skeleton with no external legs is usually defined as a graph in which one cannot isolate a self-energy by cutting two distinct lines. This definition incorporates more graphs than the ones that are enclosed in the definition of the present paper since one can then allow for $2$-insertions that are closed on each other by means of a single line.} However, this writing always involve certain terms that depend on the bare mass and the bare couplings and requires additional terms to avoid double counting. For instance, within the 2PI framework, using the notion of cycles \cite{deDominicis:1964uu}, one can show that
\beq
\ln Z=\Gamma[G]=\frac{1}{2}{\rm Tr}\,\ln G+\frac{1}{2}{\rm Tr}\,G_0^{-1}G+\Phi[G]\,,\label{eq:Z}
\eeq
where $\Phi[G]$ is the Luttinger-Ward function referred to above. The first two terms account for the overcounting of graphs that arise from the fact that there is no unique way to identify maximal $2$-insertions in $\ln Z$. Moreover the second term depends explicitely on the bare mass $m^2_{\rm b}$, so it is not possible to fully hide this bare parameter in this case (although we stress that this remaining dependence is rather trivial). Similar remarks apply to the rewriting of $\ln Z$ in terms of the three- and four-point functions leading to the so-called $3$PI and $4$PI effective actions $\Gamma[G,\Gamma^{(3)}]$ and $\Gamma[G,\Gamma^{(3)},\Gamma^{(4)}]$ which express $\ln Z$ in terms of $G\equiv[\Gamma^{(2)}]^{-1}$ and $\Gamma^{(3)}$, or $G$, $\Gamma^{(3)}$ and $\Gamma^{(4)}$ respectively. The same remarks apply to the $n$-point functions obtained by imposing a stationnarity condition to any of these functionals. For instance, the four-point function as derived from $\Gamma[G,\Gamma^{(3)},\Gamma^{(4)}]$ is given by an equation that still makes explicit reference to the quartic bare coupling. In contast, higher $n$-point functions admit a representation in which no such reference to the bare parameters appears.

\section{Conclusion}
In this article, we have studied how two arbitrary subgraphs of a given Feynman graph can overlap with each other. When restricting to 1PI subgraphs, we have shown how this allows to derive useful ``non-overlap'' theorems for the cases of $2$-, $3$- and $4$- insertions. One consequence of these is the well known skeleton expansion for vertex functions $\Gamma^{(n)}$ with $n\geq 5$ which allows one to entirely hide any reference to the bare mass, as well as the trilinear and quartic bare couplings using the two-, three- and four-point functions, and this without any over-counting correction. We have also discussed how this result can be extended to other classes of functions, in particular to iterated derivatives of the Luttinger-Ward functional.

As discussed in Ref.~\cite{Blaizot:2021ikl}, the previous results have applications in the renormalization of the 2PI effective action and the corresponding $\Phi$-derivable approximations, as well as in the construction of new truncation schemes for the functional renormalization group hierarchy but their potential range of applicability is definitely larger.

\acknowledgements{I would like to thank J.~P. Blaizot for useful discussions and collaboration on related topics and J.~P. Blaizot and D.~M. van Egmond for carefully reading the manuscript and making some useful comments.}

\appendix

\section{Number of overlap modes}
In this section, we would like to evaluate the number of overlap modes of two graphs $\bar {\cal G}_1$ and $\bar {\cal G}_2$ of $n_{\rm ext}(\bar {\cal G}_1)$ and $n_{\rm ext}(\bar {\cal G}_2)$ external lines each, when no restrictions are imposed on the graphs as in Sec.~\ref{sec:overlap}. These are determined by the constraints (\ref{eq:central}) and (\ref{eq:c1})-(\ref{eq:c3}).  To ease the reading we shall simplify the notations as $n_1\equiv n_{\rm ext}(\bar {\cal G}_1)$, $n_2\equiv n_{\rm ext}(\bar {\cal G}_2)$ and $n\equiv n_{\rm ext}(\bar {\cal G}_1\cap\bar {\cal G}_2)$. Then, the constraints rewrite
 \beq
n & \leq & n_1+n_2\,,\label{eq:c21}\\
n_{t_1}+n_{t_2} & \leq &  n\,,\label{eq:c22}\\
n_{t_1} & \geq & n-n_1\,,\label{eq:c23}\\
n_{t_2} & \geq & n-n_2\,.\label{eq:c24}
 \eeq 
 Without loss of generality, we can assume that $n_1\leq n_2$.
 
We first need to consider all possible values of $n$ compatible with (\ref{eq:c21}), that is $n_1+n_2+1$ choices in total. For each of these values, we need to choose $n_{t_1}$ and $n_{t_2}$ complying with (\ref{eq:c22})-(\ref{eq:c24}). Here, we need to consider various ranges for the values of $n$ over which (\ref{eq:c23}) and/or (\ref{eq:c24}) do not matter. More precisely, in the range $n\leq n_1$, none of these constraints matter, in the range $n_1< n\leq n_2$ only one of them matters, and in the range $n_2< n\leq n_1+n_2$, both of them. We will use the following result: given three positive integers $a,b,c$ such that $a\geq b\geq c$ and $c\leq a-b$, the conditions $n_1+n_2\leq a$, $n_1\geq b$, $n_2\geq c$ define an isosceles right triangle in the $(n_1,n_2)$-plane of sides of length $a-b-c$, corresponding to $(a-b-c+1)(a-b-c+2)/2$ points.

In the first range, the constraint (\ref{eq:c21}) gives a number of cases $(n+1)(n+2)/2$ for each value of $n$ in the range, so
\beq
\frac{1}{2}\sum_{n=0}^{n_1}(n+1)(n+2)=\frac{1}{6}(n_1+1)(n_1+2)(n_1+3)\,,
\eeq
in total. In the second range, the constraint (\ref{eq:c21}) together with (\ref{eq:c22}) or (\ref{eq:c23}) gives a number of cases $(n_1+1)(n_1+2)/2$ for each value of $n$ in the range, so
\beq
\frac{1}{2}(n_1+1)(n_1+2)(n_2-n_1)\,,
\eeq
in total. In the third range, the constraints (\ref{eq:c1})-(\ref{eq:c3}) give a number of cases $(n_1+n_2-n+1)(n_1+n_2-n+2)/2$ for each value of $n$ (one checks that one has $c\leq a-b$), so
\beq
& & \frac{1}{2}\sum_{n=n_2+1}^{n_1+n_2}\Big(n_1+n_2-n+1\Big)\Big(n_1+n_2-n+2\Big)\nonumber\\
& & \hspace{0.5cm}=\,\frac{1}{2}\sum_{n=0}^{n_1-1}(n+1)(n+2)=\frac{1}{6}n_1(n_1+1)(n_1+2)\,,
\eeq
in total. Putting all the pieces together, we arrive at
\beq
\frac{1}{6}(n_1+1)(n_1+2)(3+3n_2-n_1)\,.
\eeq
The formula for arbitrary $n_1$ and $n_2$ (that is not necessarily ordered as $n_1\leq n_2$) is obtained after replacing $n_1$ by ${\rm Min}(n_1,n_2)$ and $n_2$ by ${\rm Max}\,(n_1,n_2)$. For $n_1=n_2=0$, we find $1$. For $n_1=0$ and $n_2=1$, we find $2$. For $n_1=n_2=1$, we find $5$. This agrees with the number of overlap modes found in Tab.~\ref{table1}.

\end{document}